\RequirePackage{lineno}  

\documentclass[aps,prl,twocolumn,superscriptaddress,showpacs]{revtex4}

\usepackage{graphicx}  
\usepackage{dcolumn}   
\usepackage{bm}        
\usepackage{xspace}

\newcommand{\met}{\,/\!\!\!\!E_{T}}
\newcommand{\vecmet}{\vec{\,/\!\!\!\!E}_{T}}

\newcommand{\ttbar}{{t\bar{t}}}

\newcommand{\pairstop}{\tilde{t}_{1}\bar{\tilde{t}}_{1}}
\newcommand{\tops}{\tilde{t}_{1}}
\newcommand{\mstop}{m_{\tilde{t}_{1}}}

\newcommand{\chargino}{\tilde{\chi}_{1}^{\pm}}
\newcommand{\neutralino}{\tilde{\chi}_{1}^{0}}

\newcommand{\ppbar}{p\bar{p}}

\begin{document}

\title{ Search for Pair Production of Supersymmetric Top Quarks 
in Dilepton Events from $p\bar{p}$ Collisions
at $\sqrt{s}$ = 1.96 TeV}

\date{\today}
\affiliation{Institute of Physics, Academia Sinica, Taipei, Taiwan 11529, Republic of China} 
\affiliation{Argonne National Laboratory, Argonne, Illinois 60439} 
\affiliation{University of Athens, 157 71 Athens, Greece} 
\affiliation{Institut de Fisica d'Altes Energies, Universitat Autonoma de Barcelona, E-08193, Bellaterra (Barcelona), Spain} 
\affiliation{Baylor University, Waco, Texas  76798} 
\affiliation{Istituto Nazionale di Fisica Nucleare Bologna, $^x$University of Bologna, I-40127 Bologna, Italy} 
\affiliation{Brandeis University, Waltham, Massachusetts 02254} 
\affiliation{University of California, Davis, Davis, California  95616} 
\affiliation{University of California, Los Angeles, Los Angeles, California  90024} 
\affiliation{University of California, San Diego, La Jolla, California  92093} 
\affiliation{University of California, Santa Barbara, Santa Barbara, California 93106} 
\affiliation{Instituto de Fisica de Cantabria, CSIC-University of Cantabria, 39005 Santander, Spain} 
\affiliation{Carnegie Mellon University, Pittsburgh, PA  15213} 
\affiliation{Enrico Fermi Institute, University of Chicago, Chicago, Illinois 60637}
\affiliation{Comenius University, 842 48 Bratislava, Slovakia; Institute of Experimental Physics, 040 01 Kosice, Slovakia} 
\affiliation{Joint Institute for Nuclear Research, RU-141980 Dubna, Russia} 
\affiliation{Duke University, Durham, North Carolina  27708} 
\affiliation{Fermi National Accelerator Laboratory, Batavia, Illinois 60510} 
\affiliation{University of Florida, Gainesville, Florida  32611} 
\affiliation{Laboratori Nazionali di Frascati, Istituto Nazionale di Fisica Nucleare, I-00044 Frascati, Italy} 
\affiliation{University of Geneva, CH-1211 Geneva 4, Switzerland} 
\affiliation{Glasgow University, Glasgow G12 8QQ, United Kingdom} 
\affiliation{Harvard University, Cambridge, Massachusetts 02138} 
\affiliation{Division of High Energy Physics, Department of Physics, University of Helsinki and Helsinki Institute of Physics, FIN-00014, Helsinki, Finland} 
\affiliation{University of Illinois, Urbana, Illinois 61801} 
\affiliation{The Johns Hopkins University, Baltimore, Maryland 21218} 
\affiliation{Institut f\"{u}r Experimentelle Kernphysik, Universit\"{a}t Karlsruhe, 76128 Karlsruhe, Germany} 
\affiliation{Center for High Energy Physics: Kyungpook National University, Daegu 702-701, Korea; Seoul National University, Seoul 151-742, Korea; Sungkyunkwan University, Suwon 440-746, Korea; Korea Institute of Science and Technology Information, Daejeon, 305-806, Korea; Chonnam National University, Gwangju, 500-757, Korea} 
\affiliation{Ernest Orlando Lawrence Berkeley National Laboratory, Berkeley, California 94720} 
\affiliation{University of Liverpool, Liverpool L69 7ZE, United Kingdom} 
\affiliation{University College London, London WC1E 6BT, United Kingdom} 
\affiliation{Centro de Investigaciones Energeticas Medioambientales y Tecnologicas, E-28040 Madrid, Spain} 
\affiliation{Massachusetts Institute of Technology, Cambridge, Massachusetts  02139} 
\affiliation{Institute of Particle Physics: McGill University, Montr\'{e}al, Qu\'{e}bec, Canada H3A~2T8; Simon Fraser University, Burnaby, British Columbia, Canada V5A~1S6; University of Toronto, Toronto, Ontario, Canada M5S~1A7; and TRIUMF, Vancouver, British Columbia, Canada V6T~2A3} 
\affiliation{University of Michigan, Ann Arbor, Michigan 48109} 
\affiliation{Michigan State University, East Lansing, Michigan  48824}
\affiliation{Institution for Theoretical and Experimental Physics, ITEP, Moscow 117259, Russia} 
\affiliation{University of New Mexico, Albuquerque, New Mexico 87131} 
\affiliation{Northwestern University, Evanston, Illinois  60208} 
\affiliation{The Ohio State University, Columbus, Ohio  43210} 
\affiliation{Okayama University, Okayama 700-8530, Japan} 
\affiliation{Osaka City University, Osaka 588, Japan} 
\affiliation{University of Oxford, Oxford OX1 3RH, United Kingdom} 
\affiliation{Istituto Nazionale di Fisica Nucleare, Sezione di Padova-Trento, $^y$University of Padova, I-35131 Padova, Italy} 
\affiliation{LPNHE, Universite Pierre et Marie Curie/IN2P3-CNRS, UMR7585, Paris, F-75252 France} 
\affiliation{University of Pennsylvania, Philadelphia, Pennsylvania 19104}
\affiliation{Istituto Nazionale di Fisica Nucleare Pisa, $^z$University of Pisa, $^{aa}$University of Siena and $^{bb}$Scuola Normale Superiore, I-56127 Pisa, Italy} 
\affiliation{University of Pittsburgh, Pittsburgh, Pennsylvania 15260} 
\affiliation{Purdue University, West Lafayette, Indiana 47907} 
\affiliation{University of Rochester, Rochester, New York 14627} 
\affiliation{The Rockefeller University, New York, New York 10021} 
\affiliation{Istituto Nazionale di Fisica Nucleare, Sezione di Roma 1, $^{cc}$Sapienza Universit\`{a} di Roma, I-00185 Roma, Italy} 

\affiliation{Rutgers University, Piscataway, New Jersey 08855} 
\affiliation{Texas A\&M University, College Station, Texas 77843} 
\affiliation{Istituto Nazionale di Fisica Nucleare Trieste/Udine, I-34100 Trieste, $^{dd}$University of Trieste/Udine, I-33100 Udine, Italy} 
\affiliation{University of Tsukuba, Tsukuba, Ibaraki 305, Japan} 
\affiliation{Tufts University, Medford, Massachusetts 02155} 
\affiliation{Waseda University, Tokyo 169, Japan} 
\affiliation{Wayne State University, Detroit, Michigan  48201} 
\affiliation{University of Wisconsin, Madison, Wisconsin 53706} 
\affiliation{Yale University, New Haven, Connecticut 06520} 
\author{T.~Aaltonen}
\affiliation{Division of High Energy Physics, Department of Physics, University of Helsinki and Helsinki Institute of Physics, FIN-00014, Helsinki, Finland}
\author{J.~Adelman}
\affiliation{Enrico Fermi Institute, University of Chicago, Chicago, Illinois 60637}
\author{T.~Akimoto}
\affiliation{University of Tsukuba, Tsukuba, Ibaraki 305, Japan}
\author{B.~\'{A}lvarez~Gonz\'{a}lez$^s$}
\affiliation{Instituto de Fisica de Cantabria, CSIC-University of Cantabria, 39005 Santander, Spain}
\author{S.~Amerio$^y$}
\affiliation{Istituto Nazionale di Fisica Nucleare, Sezione di Padova-Trento, $^y$University of Padova, I-35131 Padova, Italy} 

\author{D.~Amidei}
\affiliation{University of Michigan, Ann Arbor, Michigan 48109}
\author{A.~Anastassov}
\affiliation{Northwestern University, Evanston, Illinois  60208}
\author{A.~Annovi}
\affiliation{Laboratori Nazionali di Frascati, Istituto Nazionale di Fisica Nucleare, I-00044 Frascati, Italy}
\author{J.~Antos}
\affiliation{Comenius University, 842 48 Bratislava, Slovakia; Institute of Experimental Physics, 040 01 Kosice, Slovakia}
\author{G.~Apollinari}
\affiliation{Fermi National Accelerator Laboratory, Batavia, Illinois 60510}
% ...

\collaboration{CDF Collaboration\footnote{With visitors from $^a$University of Massachusetts Amherst, Amherst, Massachusetts 01003,
$^b$Universiteit Antwerpen, B-2610 Antwerp, Belgium, 
$^c$University of Bristol, Bristol BS8 1TL, United Kingdom,
$^d$Chinese Academy of Sciences, Beijing 100864, China, 
$^e$Istituto Nazionale di Fisica Nucleare, Sezione di Cagliari, 09042 Monserrato (Cagliari), Italy,
$^f$University of California Irvine, Irvine, CA  92697, 
$^g$University of California Santa Cruz, Santa Cruz, CA  95064, 
$^h$Cornell University, Ithaca, NY  14853, 
$^i$University of Cyprus, Nicosia CY-1678, Cyprus, 
$^j$University College Dublin, Dublin 4, Ireland,
$^k$University of Edinburgh, Edinburgh EH9 3JZ, United Kingdom, 
$^l$University of Fukui, Fukui City, Fukui Prefecture, Japan 910-0017
$^m$Kinki University, Higashi-Osaka City, Japan 577-8502
$^n$Universidad Iberoamericana, Mexico D.F., Mexico,
$^o$Queen Mary, University of London, London, E1 4NS, England,
$^p$University of Manchester, Manchester M13 9PL, England, 
$^q$Nagasaki Institute of Applied Science, Nagasaki, Japan, 
$^r$University of Notre Dame, Notre Dame, IN 46556,
$^s$University de Oviedo, E-33007 Oviedo, Spain, 
$^t$Texas Tech University, Lubbock, TX  79609, 
$^u$IFIC(CSIC-Universitat de Valencia), 46071 Valencia, Spain,
$^v$University of Virginia, Charlottesville, VA  22904,
$^w$Bergische Universit\"at Wuppertal, 42097 Wuppertal, Germany,
$^{ee}$On leave from J.~Stefan Institute, Ljubljana, Slovenia, 
}}
\noaffiliation

\begin{abstract}
We present the results of a search for pair production of 
the supersymmetric partner of
the top quark (the top squark $\tilde{t}_{1}$) decaying to a
$b$-quark and a chargino $\chargino$ with a subsequent $\chargino$ 
decay  into a neutralino $\neutralino$, lepton $\ell$, and neutrino $\nu$. 
Using a data sample corresponding to~2.7~fb$^{-1}$ of integrated luminosity 
of $p\bar{p}$ collisions at $\sqrt{s} = 1.96$ TeV
collected by the CDF~II detector,  we 
reconstruct the mass of top squark candidate events and 
fit the observed mass spectrum
to a combination of standard model processes and $\pairstop$ signal.
We find no evidence for $\pairstop$ production and 
set 95\% C.L. limits 
on the masses of the top squark and the neutralino for several values 
of the chargino mass and the 
branching ratio ${\cal B}(\chargino\rightarrow\neutralino\ell^{\pm}\nu)$. 
\end{abstract}


\maketitle





Supersymmetry (SUSY)~\cite{SUSY} is a plausible extension to 
the standard model (SM) of particle physics that
naturally solves the hierarchy problem, 
predicts the unification of the gauge coupling constants,
and
provides a possible candidate for dark matter. In SUSY, a new spin-based symmetry turns a bosonic state into 
a fermionic state (and vice versa) postulating the existence of a superpartner
for each of the known fundamental particles.
To be reconciled with experimental data, SUSY must be broken,
and thus supersymmetric particles are expected to be much heavier than their SM partners. 
An exception to this might come from 
the partner of the top quark~$t$, the top squark, whose low-mass
eigenstate $\tilde{t_1}$ may be lighter than the top quark
due to the substantial top-Yukawa coupling~\cite{light_stop}.
This mass inequality  $m_{\tilde{t}_1} \lesssim m_t$ is favored in supersymmetric 
electroweak baryogenesis scenarios~\cite{EWB}.

In canonical SUSY models R-parity is conserved,
the lightest supersymmetric particle~(LSP) is the neutralino $\neutralino$, 
and top squarks $\tops$
are expected to be pair-produced via the strong interaction. 
The $\pairstop$ cross 
section depends primarily on the mass of the top squark $\mstop$, and 
at the Tevatron is expected to be an order of magnitude smaller
 than that for top quarks of the same mass~\cite{stop,ttbar}.
If the chargino~$\chargino$ is lighter than the $\tops$, the decay channel
$\tilde{t}_{1} \rightarrow b \chargino$ becomes dominant.
Subsequent chargino decays via $ \chargino \rightarrow  \neutralino \ell^{\pm} \nu$
result in experimental
event signatures with two energetic, oppositely charged leptons, 
two jets from the bottom quarks, and large imbalance in energy
from the lack of detection of the neutrinos and neutralinos.
This event signature is identical to the dilepton final state of top pair decays
($\ttbar \to W^+bW^-\bar{b} \to \ell^+\nu b\ell^{\prime-}\bar{\nu^\prime}\bar{b}$).
Therefore, an admixture of top squark events with the top dilepton events
could impact measurements of the properties of the top quark,
such as the mass value.
This search was in part 
motivated 
by apparent inconsistencies in the top 
mass measurements between different top decay channels observed
in the early CDF and D0 Run~II data~\cite{mass}.
Previous searches for top squark decays  $\tilde{t}_1\to b\chargino$~\cite{prev} 
did not exclude any region in the SUSY parameter space.

In this Letter we present the results of a search for pair production of 
scalar top quarks, each decaying as $\tilde{t}_1\to b\chargino \to b\neutralino\ell^\pm\nu$.
We analyze a data set corresponding to 2.7~fb$^{-1}$ of integrated luminosity
from $\ppbar$ collisions 
collected by the upgraded Collider Detector at Fermilab~(CDF~II)~\cite{cdf,cdf-coord}, 
and fit the data with the $\pairstop$ production hypothesis. 


We identify and record events containing $e$ or $\mu$ candidates
with large transverse momenta ($p_T \geq$ 18 GeV/$c$) 
using high-speed trigger electronics.
 The performance of the
trigger and lepton identification (ID) algorithms is described  
elsewhere~\cite{lep-id}.
We identify final state quarks as 
jets of hadrons in the calorimeter. 
Jet reconstruction employs an iterative cone-based clustering algorithm
that associates 
calorimeter energy deposits within a cone of $R
\equiv \sqrt{(\Delta\eta)^2+(\Delta\phi)^2} = 0.4$.  The energies of
reconstructed jets and the missing transverse energy ($\met$)~\cite{met_def} 
are corrected for detector non-uniformity 
and other effects~\cite{jet-energy}.
Bottom quark candidates are identified (or ``$b$-tagged")
through 
the presence within the jet cone of a displaced secondary 
vertex arising from the decay of a long-lived bottom hadron~\cite{secvtx}.

The first stage of the $\pairstop$ candidate event selection requires two leptons 
($e$ or $\mu$) with $p_T > 20 $~GeV/$c$, $|\eta| < 2.0(1.0)$ for $e(\mu)$,
at least one of which is isolated~\cite{isol}, and
$m_{\ell^+\ell^{\prime-}} > 20 $ GeV/$c^2$. 
We also require
two or more jets with $E_T > 12 $ GeV within the region $|\eta| < 2.4$, and
$\met > 20$ GeV.
For events with leptons compatible with originating from a $Z$ boson 
in the mass window from 76 GeV/c$^2$ to 106 GeV/c$^2$,
we apply a requirement on 
the missing transverse energy significance
$\met^{sig} > 4 \sqrt{\mbox{GeV}}$~\cite{met_sig}. 
Selected events are divided into two categories based on 
whether any of the jets is identified as a $b$ jet 
($b$-tagged channel) or not (non-$b$-tagged channel).
Further optimized event selection criteria  are used in
the last stage of the analysis.


The dominant SM process that contributes to the dilepton $+$ jets $+ \met$ event signature 
is $t\bar{t}$ production. Other SM processes include
 $Z/\gamma^* +$  jets, diboson, and $W +$  
jets production, where a real lepton comes from the $W$ decay and 
one of the jets is misidentified as a second lepton.
We use the \textsc{pythia} v6.216 Monte Carlo (MC) event 
generator~\cite{PYTHIA} to simulate
$\pairstop$, $t\bar{t}$, and diboson processes.
The $\pairstop$ signal is normalized according to the next-to-leading order (NLO)
theoretical cross section obtained from 
 \textsc{prospino2}~\cite{PROSPINO} 
using the CTEQ6M~\cite{cteq6m} parton density functions (PDF). 
For $t\bar{t}$ we use 
the NLO theoretical cross section value of 7.3~pb~\cite{ttbar},  corresponding to
the world average top mass of 172.5 GeV/c$^2$~\cite{top_mass}, which is 
dominated by the measurements in the lepton $+$ jets channel of $\ttbar$ decays.
Diboson processes ($WW, WZ, ZZ$) are normalized to their NLO theoretical cross-sections~\cite{diboson}. 
$Z/\gamma^*$ events with associated jets are simulated with the \textsc{alpgen} v2.13
matrix element generator~\cite{ALPGEN},
interfaced to \textsc{pythia} v6.325, and normalized to data in the $Z$-mass-peak region.
The detector response in all MC samples is modeled by a \textsc{geant}-based CDF~II detector simulation~\cite{GEANT}.
The $W +$ jets background is modeled using data
by measuring relative rates of jets being misidentified as charged leptons
in inclusive jet data samples
and applying them to data events with exactly one lepton plus jets.
We validate the background 
modeling of dilepton events by comparing the predictions with observations 
using control samples that are independent of the signal sample.
These include samples
of events with low $\met$, 
events with zero or one jet, and events with same-sign charged leptons.


To enhance the search sensitivity, we perform 
a kinematic reconstruction of events under the $\pairstop$ production
and decay hypothesis. 
We use as inputs the measured 
four-momenta of 
the two leptons and of the two largest $E_T$ jets,  and the $\vecmet$.
Due to the unknown masses of the supersymmetric 
$\chargino$ and $\neutralino$, and
because the two neutrinos and the two massive neutralinos
escape detection,
the kinematics of $\pairstop$ events is severely underconstrained.
Therefore, 
we employ the following strategy.
First, we use $m_{\chargino}$ as a fixed parameter, and 
perform the reconstruction
for different values of $m_{\chargino}$.
Second,  we treat the $\neutralino  \nu$ pair corresponding to each $\tilde{t}_1$ decay
as one ``massive particle." To compensate for non-resonant structure 
of the invariant mass of the $\neutralino  \nu$ pair we assign to this ``massive particle"
a large width. Based on studies carried out on MC samples 
for a wide range of neutralino masses ($m_{\neutralino} \approx 46 - 90$ GeV/$c^2$) 
we fix the values of $m_{\neutralino  \nu}$ and $\Gamma_{\neutralino  \nu}$
to 75 GeV/$c^2$ and 10 GeV/$c^2$, respectively.
Third, to avoid the two-fold ambiguity in assigning a $b$-jet to a lepton,
we always choose the combination that yields the smallest sum of invariant masses 
of a paired $b$-jet and lepton. 
This approach identifies the correct pairing in $\sim$90\% 
of cases in the $\tops$ mass regime considered. 

The system of kinematic equations consists of constraints imposed
on the particle masses $m_{\chargino}$, $m_{\neutralino  \nu}$, 
$m_{\tilde{t}_1} = m_{\bar{\tilde{t}_1}}$, and the requirement of 
transverse momentum conservation:
$ \vec{\met} = \vec{p}_T(\neutralino \nu)_{1} + \vec{p}_T(\neutralino \nu)_{2}.$
If the azimuthal directions $\phi_1$ and $\phi_2$ of the four-momenta of the
$(\neutralino \nu)_1$ and $(\neutralino \nu)_2$ pairs are fixed, 
 the event kinematics
(with the exception of the singular points $\phi_1 -\phi_2 = k\pi$, where $k$ is an integer)
 is constrained.
There exist four solutions
due to the two-fold ambiguities in resolving 
the $z$-components of the $(\neutralino \nu)_1$ and $(\neutralino \nu)_2$ four-momenta.
We perform a scan of the entire parameter space of azimuthal angles
wherein we repeat the reconstruction for different values of $(\phi_1,\phi_2)$, 
avoiding the aforementioned 
singular points.
The $\tops$ mass is reconstructed by minimizing the
event $\chi^2$, which takes the following form:

\begin{displaymath}
  \chi^2 =
\sum_{k= 1, 2} \left\{
\frac{ \left( m_{(\neutralino \nu)_k}^{fit} - m_{\neutralino \nu} \right)^2 }{ \Gamma^2_{\neutralino \nu} } 
+  \frac{ \left( m_{(\neutralino \nu \ell)_k}^{fit} - m_{\chargino} \right)^2 }{ \Gamma^2_{\chargino }  }
\right.  
\end{displaymath}
\begin{equation}
 \left. 
+ \frac{ \left( m_{(\neutralino \nu \ell j)_k}^{fit} -  m_{\tilde{t}_1}^{rec} \right)^2 }{ \Gamma^2_{\tilde{t}_1 }  }
\right\} 
+ \sum_{i=2\ell, 2jets}\frac{ \left( p_{T,i}^{fit} - p_{T,i}^{meas} \right) }{ \sigma_{p_{T,i}}^2 }  .
\end{equation}

\begin{figure}
  \begin{center}
\includegraphics[width=\columnwidth] 
{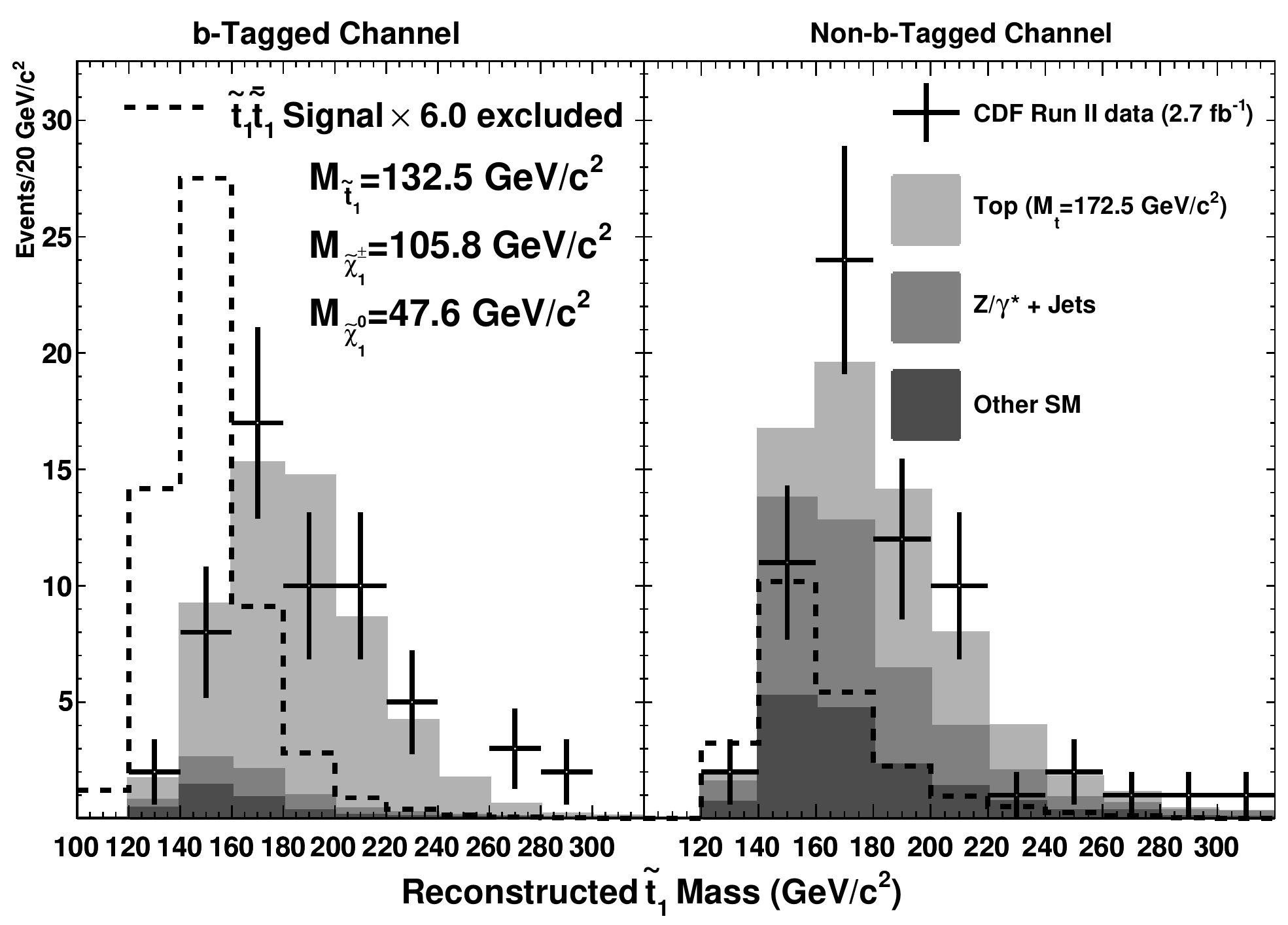}
  \end{center}
  \caption{The reconstructed $\tops$ mass distribution. 
  The dashed line represents an example of the $\pairstop$ signal distribution.
    }
  \label{fig:mass}
\end{figure}

\noindent Here we assume $\Gamma_{\chargino} \equiv $  2 GeV/$c^2$ and 
$\Gamma_{\tilde{t}_1} \equiv $  1.5 GeV/$c^2$, 
the $k$-index represents the decay products from $\tops$ or $\bar{\tops}$ respectively,
and the $m^{fit}$ are the invariant masses of the final decay products from $\tops$ decays.
We let the four-momenta of the leptons and the jets vary in the fit,
and use the \textsc{minuit} package~\cite{MINUIT} to minimize the $\chi^2$.
At each step during the minimization procedure the $\vecmet$
is re-calculated according to the values of $p_T^{fit}$ of the leptons and jets.
The longitudinal components of $(\neutralino \nu)_1$ and $(\neutralino \nu)_2$
are free parameters in the fit with starting values initialized to the values corresponding to 
solutions of the system of
kinematic equations. All four starting values are tried in the fit, but only 
the one that gives the lowest $\chi^2$ is kept.
The value $m_{\tilde{t}_1}^{rec}$ at which $\chi^2$ is minimized
yields the $\tops$ reconstructed mass 
for a given pair of the azimuthal angles $(\phi_1,\phi_2)$.
Finally, we integrate 
$m_{\tilde{t}_1}^{rec}(\phi_1,\phi_2)$ weighted by the goodness of fit term $e^{-\chi^2(\phi_1,\phi_2)}$
over $\phi_1$ and $\phi_2$ to
obtain the $\tops$ reconstructed mass 
for each event.
Running the reconstruction algorithm over simulated $\pairstop$ events yields 
a distribution with a peak near the generated $\tops$ mass, as shown in Fig.~\ref{fig:mass}, 
and provides discrimination between a $\pairstop$ signal and SM backgrounds~\cite{will}.


We perform an extended likelihood fit of the observed mass spectrum
simultaneously in the $b$-tagged and the non-$b$-tagged channels.
To quantify the level of agreement
we employ a modified frequentist method, $CL_s$~\cite{cls}, 
based on a log-likelihood ratio test statistic, which involves 
computing $p$-values 
under the hypothesis of the SM background only
and the hypothesis of signal plus background.
The systematic uncertainties for both signal and background, described below,
enter the fit as Gaussian-constrained nuisance parameters.  
The uncertainties due to kinematic mis-modeling are taken into account by allowing 
the reconstructed mass distributions to change 
according to the values of the nuisance parameters~\cite{morph}.


Imperfect knowledge of various experimental and theoretical 
parameters leads to systematic 
uncertainties that degrade our sensitivity to a $\pairstop$ signal. 
The dominant systematic effect is due to the uncertainties in the NLO 
theoretical cross sections for $\pairstop$ and $t\bar{t}$ production.
These uncertainties come from two sources: the renormalization and factorization scale 
(11\% and 7\% for $\pairstop$ and $t\bar{t}$,  respectively)
and PDFs (14\% and 7\%)~\cite{stop,ttbar}.
We assume that the scale uncertainty is uncorrelated 
for $\pairstop$ and $t\bar{t}$ processes, 
while the PDF uncertainty is fully correlated.
The theoretical uncertainty of the diboson cross sections is 10\%~\cite{diboson}, 
and assumed to be uncorrelated with other systematic uncertainties. 
The experimental uncertainties applied to MC-based background estimates 
include those due to jet energy scale~(3\%), 
$b$-tagging probability~(5\%), lepton ID and trigger efficiencies~(1\% per lepton), 
initial and final state radiation~(2\%), and the integrated luminosity~(6\%).
The uncertainty on $W$+jets is dominated by the uncertainties in the rate to misidentify jets
as leptons~(30\%), while the uncertainty on $Z/\gamma^*+$ jets comes from MC mis-modeling of the
high-$\met$ tail, jet multiplicity distribution and $Z/\gamma^*+$ heavy-flavor contribution~(16\%). 

\begin{table}
  \begin{tabular}{ccccccc}
    \multicolumn{7}{c}{Events per 2.7 fb$^{-1}$ in the signal region.}
    \\ \hline
    \hline  
                           &       Top &  $Z/\gamma^*$+jets     &   Diboson   &    $W$+jets   &  Total    &   Data \\ \hline
     $b$-tag  &      49.0$\pm$6.9  &   4.0$\pm$0.4 & 0.5$\pm$0.1  &  2.8$\pm$0.9 & 56.4$\pm$7.2 & 57 \\
    no tag  & 25.2$\pm$3.3 & 25.0$\pm$5.6 &  6.0$\pm$1.3 & 9.8$\pm$2.9 & 65.9$\pm$9.8 & 65 \\
    \hline\hline
       \end{tabular}  
 \caption{The expected event yields from SM processes
  with the total uncertainties and the observed numbers of events 
  in the signal region. 
              }
  \label{tab:result}
\end{table}

\begin{figure}
  \begin{center}
  \includegraphics[width=\columnwidth] 
  {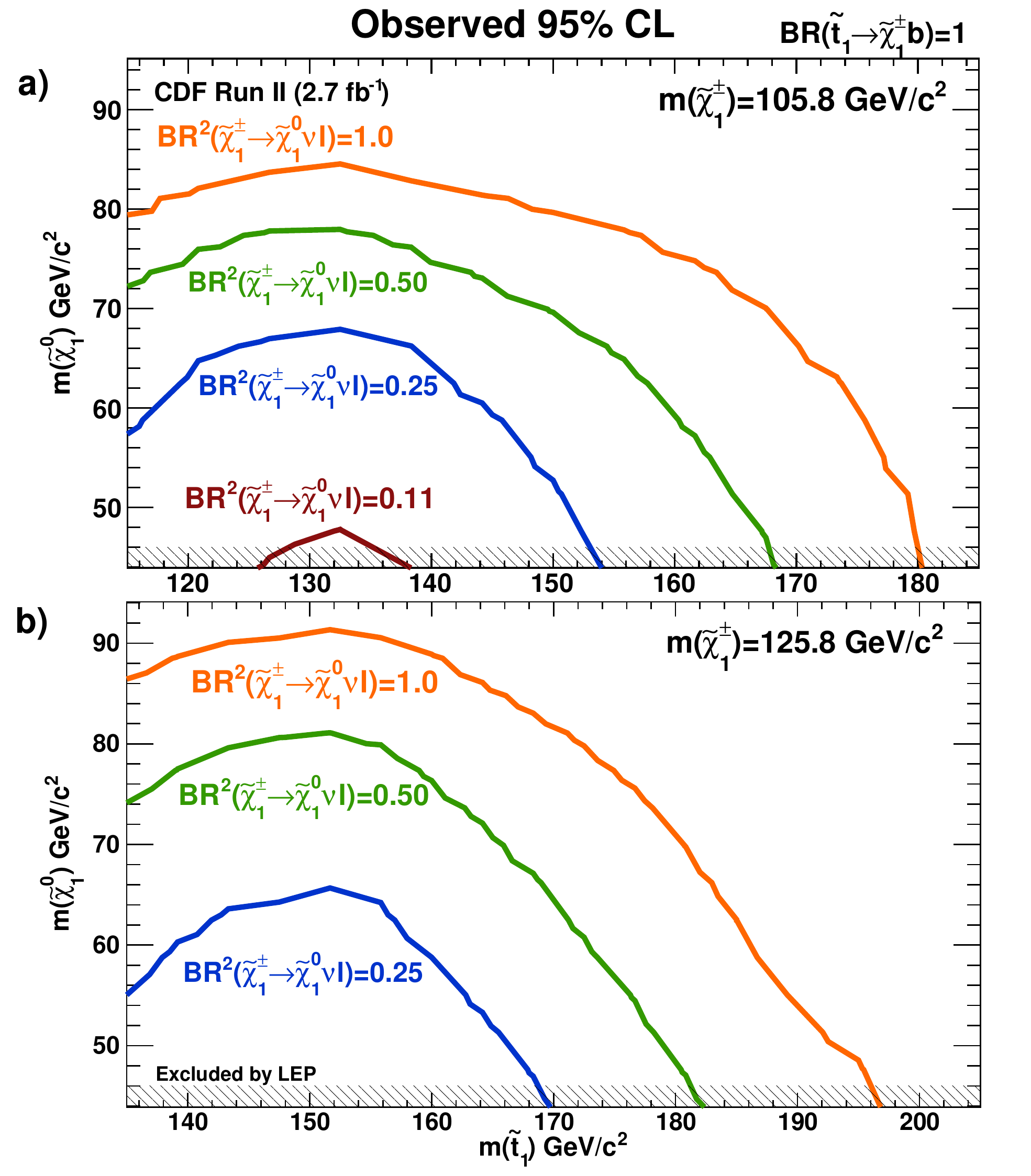}
  \end{center}
  \caption{  The observed 95\% C.L.\ exclusion regions in the 
                   $m_{\neutralino}$  and $m_{\tilde{t}_1}$  mass plane for several values of
                 ${\cal B}$($\chargino\rightarrow\neutralino\ell^{\pm}\nu$)
                 and  $m_{\chargino} $.
		The excluded region corresponds to the area below the lines.
      		Universality of $e$, $\mu$, and $\tau$
                  in the $\chargino$ decays is assumed.
                }
  \label{fig:limit125}
\end{figure}

Prior to looking at data in the signal sample we study the sensitivity of the search, 
taking into account all systematic effects, 
for various event selection criteria
imposed separately for the $b$-tagged and the non-$b$-tagged channels. 
An algorithm based on biological 
evolution (a so-called genetic algorithm)~\cite{genetic}
is employed to determine 
the most sensitive selection criteria.
Requirements yielding poorer expected 95\% C.L. limit
are culled, while those improving the limit are bred together 
until reaching a plateau. 
This procedure optimizes the event selection criteria directly to produce the best expected 95\% C.L.\ limit
in the no-signal hypothesis.

In the $b$-tagged (non-$b$-tagged) channel the optimization procedure yields 
the following event selection 
criteria~\cite{optim}:
the leading jet $E_T$ is required to be 
greater than 15 (20) GeV, and the sub-leading jet $E_T$ must be greater 
than 12 (20) GeV. 
In both channels we require $\met > 20 $ GeV, 
while this requirement is tightened to 50 GeV in the non-$b$-tagged channel if there is a lepton 
or jet within an azimuthal angle of $20^\circ$ from the $\vec{\met}$ direction.
Due to the fact that the $\tops$ is a scalar particle, and the top quark is a fermion, 
the angular distributions of their final decay products are very distinct. Therefore
we implement an additional topological cut 
in both the $b$-tagged and non-$b$-tagged channels
to suppress $\ttbar$ events: 
\begin{equation}
\sum {p_T} < \left(\frac{\Delta\phi_{jj} \times \Delta\phi_{\ell\ell}}{\pi^2} \times 325 + 215\, \right)\mbox{GeV/c},
\end{equation}

\noindent where $\sum {p_T}$ is the scalar sum of transverse momenta of the leptons, jets and the $\met$, 
the $\Delta\phi_{jj}$ and $\Delta\phi_{\ell\ell}$ are the azimuthal angles between the jets and leptons,
respectively, and the numerical values are the result of the optimization procedure.
This requirement rejects about 50\% of $\ttbar$ events and only about 10\% of $\pairstop$ events.


After applying these event selection requirements we obtain the numbers of predicted 
and observed events listed in Table~\ref{tab:result}.
The data distributions of the reconstructed $\tops$ mass in both channels 
are shown in Fig.~\ref{fig:mass}, 
together with the expectations from 
SM processes and an example of $\pairstop$ signal.
The data are consistent with the SM alone and 
there is no evidence of $\pairstop$ production.
We use these results to calculate the  95\% C.L.  exclusion limits  
in the $m_{\neutralino} $ vs $m_{\tilde{t}_1}$ plane for several values 
of the branching ratio ${\cal B}(\chargino\rightarrow\neutralino\ell^{\pm}\nu)$
and $m_{\chargino}$, 
assuming equal branching ratios into different lepton flavors
and ${\cal B}(\tilde{t}_1 \to \chargino b) = 100\%$.
The limits for two different values of 
chargino mass
 are presented in Fig.~\ref{fig:limit125}.
For a given branching ratio of the pair of top squarks decaying into leptons, 
equal to 
${\cal B}^2(\chargino\rightarrow\neutralino\ell^{\pm}\nu)$,
we exclude the top squark and neutralino masses below the respective curve shown in the plot.
The values ${\cal B}^2(\chargino\rightarrow\neutralino\ell^{\pm}\nu)$ are expected to range from almost $100\%$, 
corresponding to the scenario with 
light sleptons and sneutrinos  ($m_{\tilde{\ell}}, m_{\tilde{\nu}} \gtrsim m_{\tilde{t_1}}$), 
where the leptonic decay of the chargino goes mostly through virtual sleptons and sneutrinos,
 down to $11\%$, where sleptons and sneutrinos are heavy ($m_{\tilde{\ell}}, m_{\tilde{\nu}} \gg m_{W}$)
 and the chargino decay through a virtual $W$ is dominant.
For the scenario corresponding to the case in which the masses 
of the chargino and neutralino are near the current lower LEP exclusion limits,
$m_{\chargino} = 105.8 \,\mbox{GeV}/c^2$,
$m_{\neutralino} = 47.6 \,\mbox{GeV}/c^2$~\cite{LEP_SUSY}, 
we exclude a top squark with masses between 128 and 135 GeV/c$^2$ 
at 95\% C.L. independent of the value of 
${\cal B}^2(\chargino\rightarrow\neutralino\ell^{\pm}\nu)$.
The limits obtained are applicable to any R-parity conserving SUSY scenario where
the neutralino is the LSP and  the $\tops$decays exclusively into $\chargino b$, and are the first 
lower limits on $\tops$ mass in this mode.


In conclusion, we have presented the results of a search for pair production of 
supersymmetric top quarks decaying via $\tilde{t}_1 \to b\chargino \to b\neutralino\ell^\pm\nu$
using a data sample corresponding to 2.7 fb$^{-1}$ of integrated luminosity in 1.96 TeV 
$p\bar{p}$ collisions.  Our fit to the observed
$m^{rec}_{\tilde{t}_1}$ distribution reveals no evidence for 
$\pairstop$ production, and we place the world's first limits 
 on the masses of $\tilde{t}_1$ and $\neutralino$ for several values of $m_{\chargino}$
 and branching ratio of
 ${\cal B}(\chargino\rightarrow\neutralino\ell^{\pm}\nu)$ in this mode.


We thank the Fermilab staff and the technical staffs of the participating institutions for their vital contributions. This work was supported by the U.S. Department of Energy and National Science Foundation; the Italian Istituto Nazionale di Fisica Nucleare; the Ministry of Education, Culture, Sports, Science and Technology of Japan; the Natural Sciences and Engineering Research Council of Canada; the National Science Council of the Republic of China; the Swiss National Science Foundation; the A.P. Sloan Foundation; the Bundesministerium f\"ur Bildung und Forschung, Germany; the World Class University Program, the National Research Foundation of Korea; the Science and Technology Facilities Council and the Royal Society, UK; the Institut National de Physique Nucleaire et Physique des Particules/CNRS; the Russian Foundation for Basic Research; the Ministerio de Ciencia e Innovaci\'{o}n, and Programa Consolider-Ingenio 2010, Spain; the Slovak R\&D Agency; and the Academy of Finland.



\begin{thebibliography}{99}
\bibitem{cdf-coord} CDF uses a cylindrical coordinate system with the $z$ axis along the proton beam axis. 
 				$\theta$ is the polar angle relative to the proton beam direction,
				and 	$\phi$ is the azimuthal angle.
				Pseudorapidity is  defined as $\eta \equiv -\ln(\tan\frac{\theta}{2})$, 
				 while transverse momenta and energies of particles are defined as 
                                      $p_T = |p|\sin\theta$ and $E_T = E\sin\theta$ respectively.
\bibitem{met_def} The missing transverse energy is defined as $\met = |\vecmet| = | - \sum_i E_T^i \vec{n_i}|$ 
where $\vec{n}_i$ is a unit vector in the plane transverse to the beam direction and that points
from the event vertex to the $i$-th calorimeter tower. The missing transverse energy is corrected 
for the escaping muon momenta when muon candidates are present.
\bibitem{isol} The lepton is defined as isolated, if the energy deposit within 
$R=0.4$ cone of the lepton momentum $p_T$ is less than 10\% of the lepton $p_T$.
\bibitem{met_sig} The missing transverse energy significance is defined as $\met^{sig} = 
\met / \sqrt{\sum E_T}$, where $\sum E_T$ is the scalar transverse energy sum over all calorimeter 
towers.  $\sum E_T$ is corrected 
in an identical manner to the $\met$.
\bibitem{optim} Event selection criteria are optimized for the signal, assuming
${\cal B}^2(\chargino\rightarrow\neutralino\ell^{\pm}\nu) = 25\%$, 
and found to be invariant to assumptions on masses of SUSY particles.
\end{thebibliography}
\end{document}